\documentclass[conference]{IEEEtran}
\IEEEoverridecommandlockouts
\usepackage{cite}
\usepackage{mathptmx}
\usepackage{amsmath,amssymb,amsfonts}
\usepackage{graphicx}
\usepackage{textcomp}
\usepackage{xcolor}
\usepackage{inconsolata}

\usepackage{comment}
\usepackage{algorithmic}
\usepackage[ruled]{algorithm2e}
\specialcomment{ali}{\begingroup\sffamily\color{green}}{\endgroup}
\specialcomment{pai}{\begingroup\small\sffamily\color{purple}}{\endgroup}
\newcommand{\red}[1]{\textcolor{red}{#1}}

\newcommand{\Newport}{Newport}
\newcommand{\id}[1]{\ensuremath{\mathit{#1}}}
\newcommand{\Sec}[1]{Section~\ref{#1}}
\newcommand{\Fig}[1]{Fig.~\ref{#1}}

\def\BibTeX{{\rm B\kern-.05em{\sc i\kern-.025em b}\kern-.08em
    T\kern-.1667em\lower.7ex\hbox{E}\kern-.125emX}}
\begin{document}

\title{STANNIS: Low-Power Acceleration of Deep Neural Network Training Using Computational Storage Devices}

\author{
\IEEEauthorblockN{1\textsuperscript{st} Ali HeydariGorji}
\IEEEauthorblockA{\textit{EECS Department} \\
\textit{University of California, Irvine}\\
heydari@uci.edu
}
\and
\IEEEauthorblockN{2\textsuperscript{nd} Mahdi Torabzadehkashi}
\IEEEauthorblockA{\textit{NGD Systems Inc.}\\
mahdi.torabzadeh@ngdsystems.com
}
\and
\IEEEauthorblockN{3\textsuperscript{rd} Siavash Rezaei}
\IEEEauthorblockA{\textit{Computer Science Department} \\
\textit{University of California, Irvine}\\
siavashr@uci.edu
}
\and
\IEEEauthorblockN{4\textsuperscript{th} Hossein Bobarshad}
\IEEEauthorblockA{\textit{NGD Systems Inc.}\\
hossein.bobarshad@ngdsystems.com
}
\and
\IEEEauthorblockN{5\textsuperscript{th} Vladimir Alves}
\IEEEauthorblockA{\textit{NGD Systems Inc.}\\
vladimir.alves@ngdsystems.com
}
\and
\IEEEauthorblockN{6\textsuperscript{th} Pai H. Chou}
\IEEEauthorblockA{\textit{EECS Department} \\
\textit{University of California, Irvine}\\
phchou@uci.edu
}
}

\maketitle
\begin{abstract}
This paper proposes a framework for distributed, in-storage training of neural networks on clusters of computational storage devices. Such devices not only contain hardware accelerators but also eliminate data movement between the host and storage, resulting in both improved performance and power savings. More importantly, this in-storage processing style of training ensures that private data never leaves the storage while fully controlling the sharing of public data. Experimental results show up to 2.7x speedup and 69\% reduction in energy consumption and no significant loss in accuracy.

\end{abstract}

\section{Introduction}

In the past decade, the amount of data generated has grown at an exponential rate. In 2018, over 2.5 quintillion bytes of data were generated every day \cite{ref_int_domo}. That equals 2.5 million terabyte hard disk drives (HDD) every single day. The emergence of the Internet of Things (IoT) is further accelerating this growth of data generation. Data is not meaningful unless processed; moreover, data often need to be processed in real-time or near real-time. 
For example, in self-driving cars, a large amount of data needs to be processed quickly and locally. The rate of data generation is very high, and the required processing is very heavy. Thus, they can create a backlog of unprocessed data, which prevents the system from making a proper decision. Although new metrics such as Age of Information (AoI) can help preserve the freshness of data by discarding the old data and replacing it with new, valid data \cite{ref_int_AoI}, the problem with processing still persists.

With the rise in the use of machine learning algorithms, many problems can be solved at the cost of high computation. Up until recent years, the conventional method to process the massive volume of data was to send it to data centers, process, and resend the results back to the client. This method is associated with common problems such as latency, identity management, and data protection\cite{ref_int_cloud}. With the considerable progress in the design of low-power processors such as ARM\textsuperscript{\textregistered} A series, the edge computing has emerged as an alternative solution. Although edge computing can be beneficial for small size applications, data centers remain the backbone of heavy computation tasks, including the training of neural networks. Based on the size of the network and the number of samples, the training procedure can take weeks, even on the most powerful processing machines currently existed. Distribution of the training can significantly reduce the overall time, but obstacles to effective distribution include synchronization of the models, data transfer, and power dissipation.

In this paper, we propose a distributed in-storage processing (ISP) framework to address the issues above. The ISP technology is to process data inside the storage devices without sending the data to the host. The ISP-enabled storage devices, known as Computational Storage Devices (CSD), are equipped with an internal processing engine that is capable of processing data in-place. This technology is studied in the literature and has been shown to be beneficial for big data and HPC applications \cite{Torabzadeh_bigdata,jun2015bluedbm,torabzadeh_HPCC,gu2016biscuit}.
This technology reduces the data movements, and consequently, it improves both performance and energy consumption.
Additionally, in cases where the applications are connected to the storage devices via a network, data should move through this network to be processed. This long-range data movement raises some concerns regarding the security and the privacy of data and makes it vulnerable to privacy attacks. The ISP technology completely solves this problem by avoiding the data movement.
The overall contributions of this paper are as follows:

\begin{itemize}
\item a novel low-power CSD named {\Newport} with augmented processing power;
\item a framework named Stannis to efficiently parallelize training tasks on clusters of CSDs;
\item a tuning algorithm to maximize the utilization of a heterogeneous system;
\item protecting privacy with a hybrid of private and public data in a storage system.
\end{itemize}
The rest of the paper is structured as follows. \Sec{sec:bg-related} reviews related work on neural network training and computational storage devices.  \Sec{sec:tech} introduces the {\Newport} hardware and software stack, followed by our distribution framework named  Stannis. \Sec{sec:exp} presents our experimental results.

\section{Background and Related Work}
\label{sec:bg-related}
This section provides a background on the background on deep neural networks and their efficient processing on novel parallel archiectures. 

\subsection{Deep Neural Networks}

Deep learning has achieved tremendous success and has become one of the most powerful modeling tools combined with neural network. The idea behind neural network is the imitation of human neurons and is developed based on the perceptron model. Deep learning stacks multiple layers of neural network, in which the output of the current layer is fed as the input to the next layer. As an example, the supervised learning trains the neural network by taking the training dataset as input and comparing the output with the corresponding output label. After several training iterations, the parameters in the neural network are updated and able to do the prediction using the same type of input. Thanks to the idea of chain rule, back propagation is developed to train neural network more efficiently. As the name of the algorithm suggests, the update of parameters starts from the last layer and propagate back to the previous layer based on the comparison between the output and labels. In order to optimize the training process, Stochastic Gradient Descent (SGD) is used to reduce the error. Many machine learning libraries have been developed for model developing using neural networks, including Tensorflow, Theano, and PyTorch. However, due to the size and complexity of neural networks and the huge amount of training data, the training period has grown drastically from hours to days and weeks.

\subsection{Parallelization of Training}

A practical approach to reducing the training time is to parallelize it on multiple devices. Two common methods for parallelization are the model parallel and data-parallel. In the former, each processing node is responsible for training a part of the neural network; in the latter, all processing nodes have a replica of the entire network and update it locally, but since different copies of the network get different updates, a synchronization method is needed. The initial version of Tensorflow, a well known library for machine learning applications, uses a parameter server approach to synchronize worker nodes\cite{ref_bck_tensorflow}. Each worker sends parameter updates to the parameter server where all updates are accumulated, averaged, and then sent back to the workers.
This method can cause either computation or communication congestion on the server-side. Pytorch\cite{ref_bck_pytorch} uses a similar approach with a more enhanced allreduce algorithm \cite{ref_bck_allreduce}.
In 2017, Horovod \cite{ref_bck_horovod} for distributed training of neural networks has drawn attention with its superior scalability compared to Tensorflow's standard distribution framework.
Horovod was developed based on Nvidia's NCCL ring-allreduce algorithm\cite{ref_bck_ringallreduce}, where each node communicates only with two peer nodes and pass updates along in a circular fashion. This method is bandwidth optimal, and each node's communication bandwidth is independent of the number of the nodes. However, where Horovod falls short is the support for heterogeneous processors, a problem we will address in \Sec{sec:tech}.

\subsection{Hardware Accelerators}

Another way to decrease the overall time is to use more powerful devices. Due to its nature of parallel architecture, GPUs are much faster than regular CPUs in floating-point operations that commonly occur in training. For instance, Nvidia Tesla V100 can run a maximum of 100 TFlops whereas an Intel\textsuperscript{\textregistered} Core™ i9-7980XE can output a maximum of 1.3 TFlops. In 2016, Google unveiled its machine-learning chip called Tensor Processing Unit (TPU), a massively paralleled Application Specific Integrated Circuit (ASIC) for tensor operations that can deliver up to 420 TFlops in its latest version \cite{ref_bck_benchmark}. 
However, the downside of both GPU and TPU is that they are application-specific processors and are not as good as CPU for general operations.
Field programmable gate arrays (FPGAs) are also widely used in accelerating different applications, however, they have a long design time and need to be reconfigured for executing different operations\cite{rezaei_ICCD}.

\subsection{In-Storage Processing}

Before the era of NAND flash memories, the I/O speed on hard disk drives (HDD) was the main obstacle in accelerating the operations. Although NAND flash memories are faster than their magnetic counterparts, moving data from the storage to the DRAM for processing still remains as the bottleneck. ISP is an innovative technology that addresses this shortcoming by moving computation to the data instead of moving data to the computation. ISP allows tasks to be executed in place, i.e., in the storage devices, with no need to move huge volumes of data. Based on the processing units, the available CSDs can be categorized into two main groups: 1) CSDs that exploit the processing unit dedicated to SSD controller for ISP \cite{kim2016storage, park2016storage}, and 2) CSDs with a dedicated ISP engine \cite{torabzadeh_PDP,jun2015bluedbm, gu2016biscuit}.
In the first approach, despite the energy efficiency of exploiting the SSD controller's processing units for the purpose of ISP, the gained processing power is rather limited due to their relatively low performance; moreover, doing so can also negatively impact the performance of the SSD.
In the second approach, dedicated processing units such as FPGAs \cite{kim2016storage} or embedded processors \cite{ref_rltd_compstor} overcome the limited processing power by augmenting the storage units with more computing resources. Although FPGAs can be power efficient with a great performance improvement potential, they suffer from several limitations that make them unsuitable for ISP purpose, including
1) they require an RTL implementation of the tasks, whose high-performance realization can be very challenging;
2) reconfiguration of FPGAs is needed for different tasks;
3) due to the lack of a file system structure, data cannot be accessed and used as a file concept. It has been shown that deploying a dedicated processor is a viable solution for ISPs \cite{torabzadeh_PDP}. In this paper, we propose a dedicated processor-based CSD and show the benefits of using it in the distributed training phase of learning algorithms.

\section{CSD architecture}
\label{sec:tech}

In-storage processing technology aims at augmenting storage devices with efficient processing engines, enabling them to process data locally without transferring it to a host system. Despite the simplicity of the concept, many of the previously proposed CSD architectures fall short of providing compelling flexibility and efficiency simultaneously.

\begin{figure}
\centerline{\includegraphics[scale=0.45]{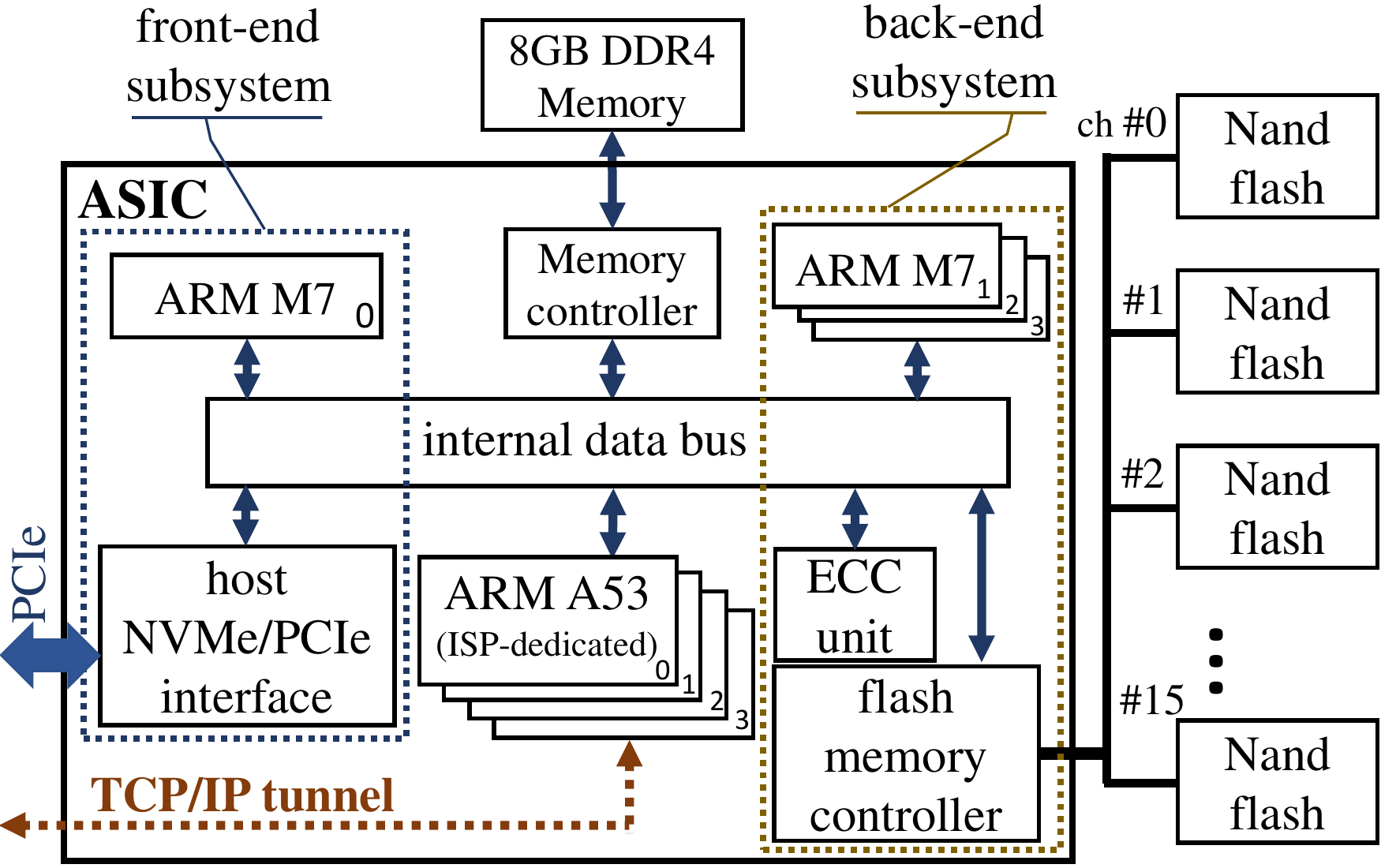}}
\caption{{\Newport} architecture}
\label{Newport_hardware_fig}
\end{figure}

In this research, we introduce {\Newport}, an ASIC-based CSD architecture that provides a flexible and capable ISP environment for offloading and running user applications without modification. {\Newport} is equipped with a full-fledged OS, which can support a wide range of programming models and languages. {\Newport} is designed to embrace the ISP technology natively, which means its controller is composed of a dedicated ISP processing engine together with other conventional modules, all implemented in a single ASIC chip. \Fig{Newport_hardware_fig} shows the hardware architecture of the {\Newport} CSD, whose internal modules are categorized into three subsystems: front-end subsystem (FE), ISP-dedicated processing engine, and back-end subsystem (BE).
{\Newport} includes four ARM\textsuperscript{\textregistered} M7 processors for running the conventional tasks of the storage device. The FE subsystem uses one of the four ARM\textsuperscript{\textregistered} M7 processors and a host NVMe interface module. The interface module receives and depacketizes the NVMe commands, and the ARM\textsuperscript{\textregistered} M7 processor checks the integrity of the commands and interprets them for the BE subsystem.

The CSD controller is connected to multiple NAND flash chips organized in 16 flash channels, which can do I/O operations in parallel. The BE subsystem manages all the flash channels and performs the I/O operations on the channels. The BE contains a flash memory controller unit, three ARM\textsuperscript{\textregistered} M7 processors, and an error correction code (ECC) unit for restoring the original data in case of bit errors from flash access. The BE also handles flash management tasks, including wear leveling, garbage collection, and flash memory logical-to-physical address translation layer (FTL). Besides these two subsystems, {\Newport} is equipped with a unique and dedicated ISP engine composed of a quad-core ARM\textsuperscript{\textregistered} Cortex A-53 processors and a complete software stack. The quad-core processor has access to an 8GB DRAM, which is shared among the processing units inside the CSD controller. The Software stack makes it possible to send ISP commands from the host to the {\Newport} CSD and sending the ISP results back to the host.

All the subsystems in the {\Newport} architecture are connected by a data bus, as shown in \Fig{Newport_hardware_fig}. Both the FE and ISP subsystems send flash read/write commands to BE, which is responsible for communicating with the flash memory chips to complete the flash I/O operations. In other words, while the data that is transferred to the host needs to go through the FE subsystem and a complex NVMe over PCIe link, the ISP engine bypasses the FE as well as the power-consuming NVMe link to read the data. Thus, the ISP subsystem has efficient and high-speed access to the data stored in the flash memory. 

\begin{figure}
\centerline{\includegraphics[scale=0.4]{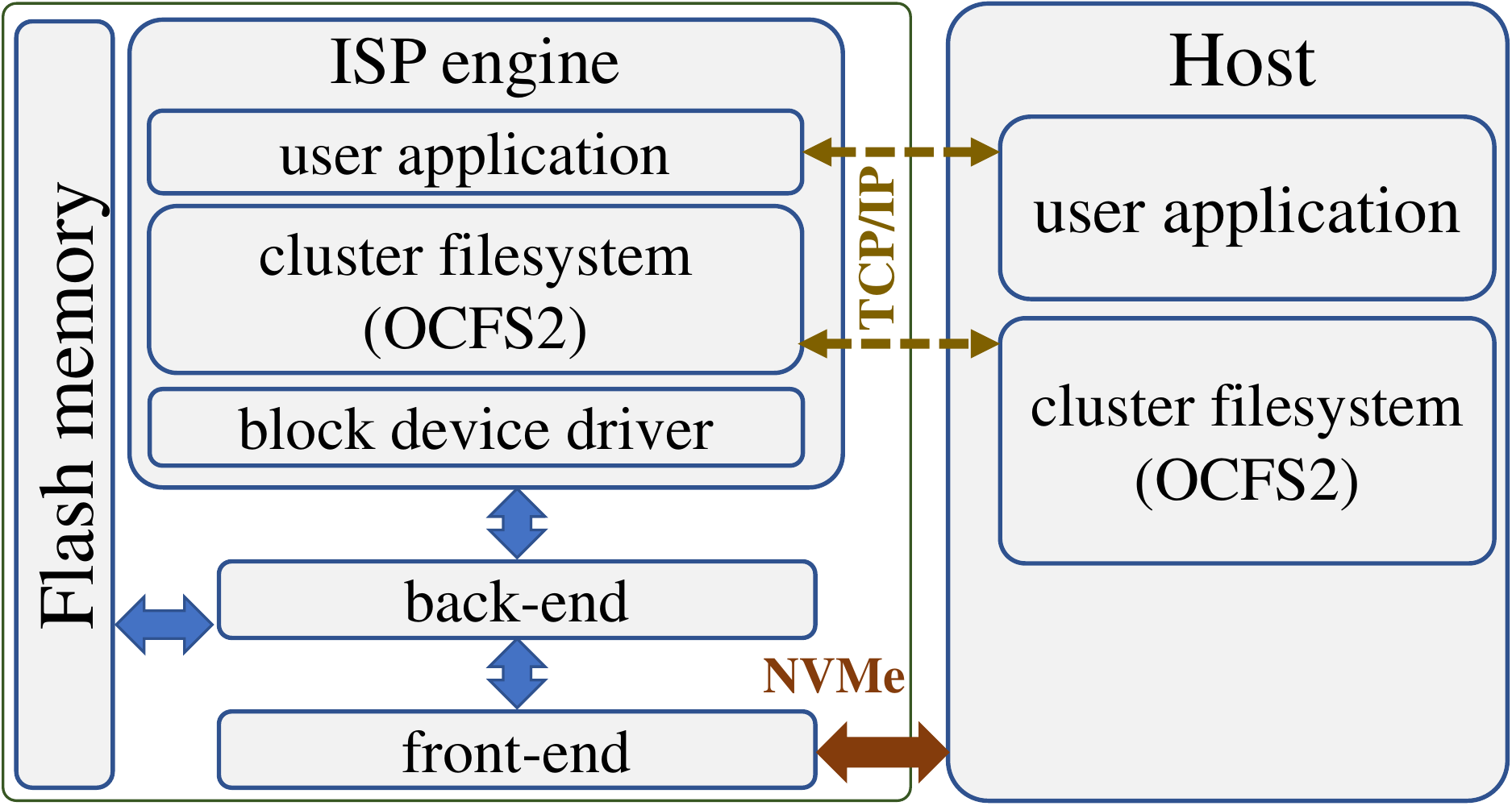}}
\caption{{\Newport} software stack}
\label{Newport_software_fig}
\end{figure}

\Fig{Newport_software_fig} shows {\Newport} software stack architecture. The ISP engine runs a full-fledged Linux operating system (OS) both for running a wide spectrum of user applications in-place as well as developing the software stack.
To provide filesystem-level access for user applications that run on the ISP engine, we have developed a \emph{block device driver}, which makes an abstraction layer for flash I/O operations, so the OS on the ISP engine can mount the flash memory as a block device. 

For communication purposes, a TCP/IP over PCIe tunnel is provided, which embeds TCP/IP packets inside the PCIe packets. Using this tunnel, user applications running on the host can communicate with the applications running on {\Newport}. The TCP/IP tunnel is composed of a process running on the FE to transfer data from the ISP engine to the host and vice versa, a process running on the host which handles TCP/IP packetization on the host, and a similar process running on the {\Newport} CSD to handle the packetization. These three processes work together to provide a TCP/IP connection between the ISP engine and the host. In the case of attaching multiple {\Newport} CSDs to a host, the TCP/IP over PCIe tunnel spans over all the CSDs to provide a network covering all the CSDs and the host. Using the {block device driver} and the TCP/IP tunnel, user applications running on {\Newport} have efficient and high-speed access to flash data, and can also communicate with the applications running on the host or other CSDs.

\begin{figure}
\centerline{\includegraphics[scale=0.3]{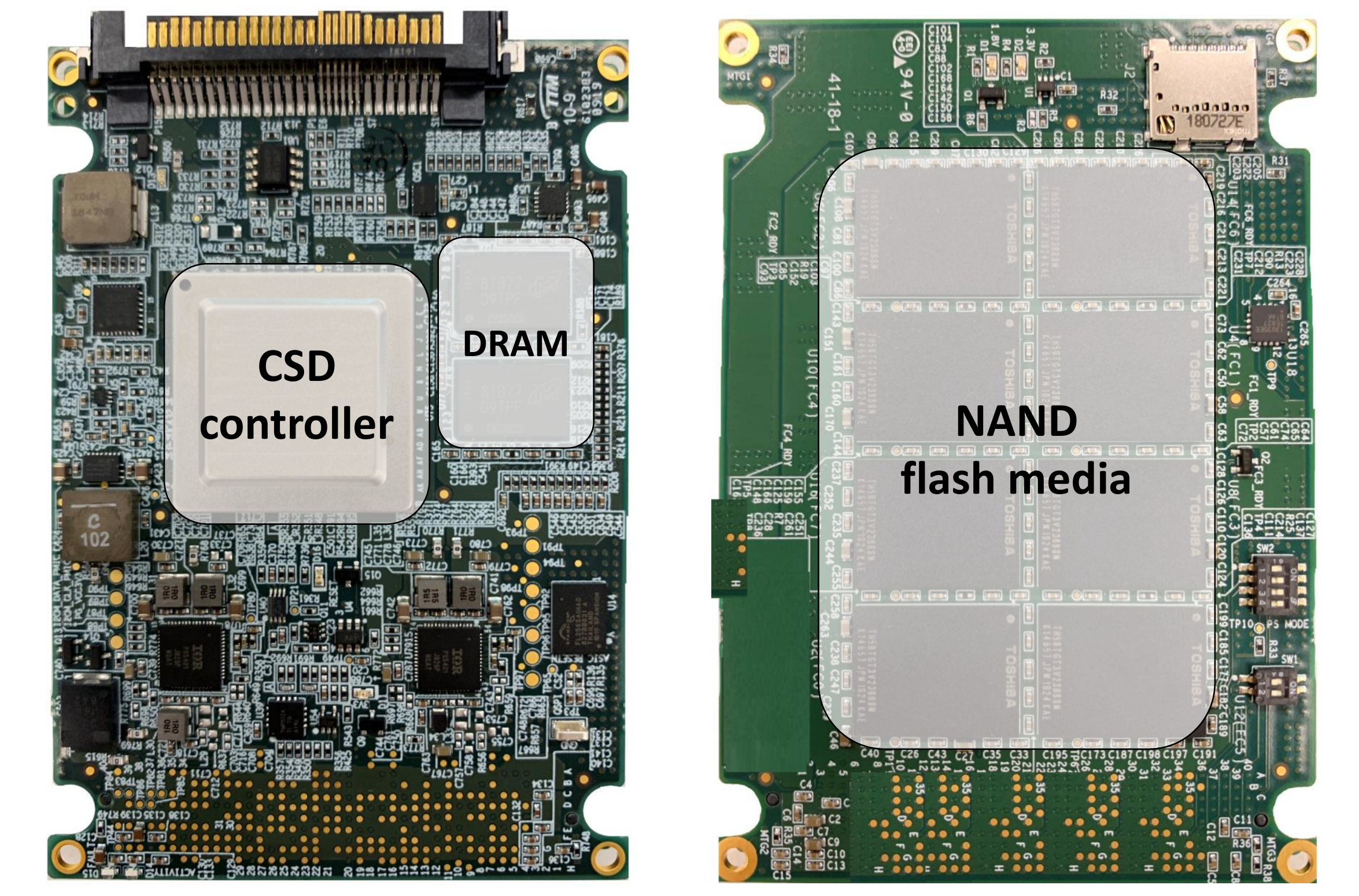}}
\caption{{\Newport} Prototype}
\label{newport_prototype}
\end{figure}

Since both the host and the ISP engine have concurrent access to the flash memory at the filesystem level, a synchronization mechanism should maintain the integrity of the filesystem metadata. To address this issue, we have ported Oracle Cluster FileSystem version 2 (OCFS2) on both the host and {\Newport} CSD. OCFS2 has two agents that run on the host and the {\Newport} CSD. These two agents communicate with each other through the TCP/IP tunnel to synchronize the filesystem metadata. We have prototyped a fully functional {\Newport} CSD to show the feasibility of the proposed design and also to run experiments on a platform equipped with multiple {\Newport} CSDs. \Fig{newport_prototype} demonstrates the {\Newport} prototype.

\section{Software Framework}

System for TrAining of Neural Networks In Storage (\textit{Stannis}) is a framework for the distribution of neural network training on homogeneous and heterogeneous systems. It overcomes Horovod's inability to work efficiently on hetereogeneous systems by workload scaling with consideration for transmission delay and data privacy.
 
Horovod shows great speedup on homogeneous systems, but on a heterogeneous system, the slowest processor becomes the bottleneck due to required synchronization in training. That is, a faster processing engine must wait for the slower ones. To address this problem, we try to equalize the processing time by setting different batch sizes for each processing engine, so that all processors wait for the least amount of time. In other words, the slower processor gets smaller batch size and thus, finishes the epoch in the same elapsed time as the more powerful processors. 

Algorithm~\ref{alg:alg_1} shows the pseudo code for the proposed approach. It starts by running a series of benchmarks on all processing engines to assess their processing speed and find the optimal batch size on each processor. Based on the results, the best batch size is selected for the slowest engine.
Having the batch size for one device, we calculate the time it takes to finish one batch and find the best batch size on the other processing engines that can deliver almost the same elapsed time. We increase the batch size by a fraction ($\frac{1}{C}$) of the difference in time for the two processor, up until getting similar times. $C$ is a constant that determines how much to adjust the batch size after each test. Larger C means more fine grained batch size update.
We also consider the slowdown that occurs due to the synchronization process and allocates a ($\frac{time}{E}$) margin to the final time. We determined $E$ by observing the slowdown pattern for the first number of processing nodes added to the system in our initial tests and defined it to give a fixed 20\% margin to the results.
Stannis also considers the access permissions of the data and assigns the private data to local ISP engine while sharing the public data with the ISP engine and the host processor. This method eliminates the transmission of private data over the network and to the host and increases the data protection level.
\begin{figure}
\centerline{\includegraphics[scale=0.4]{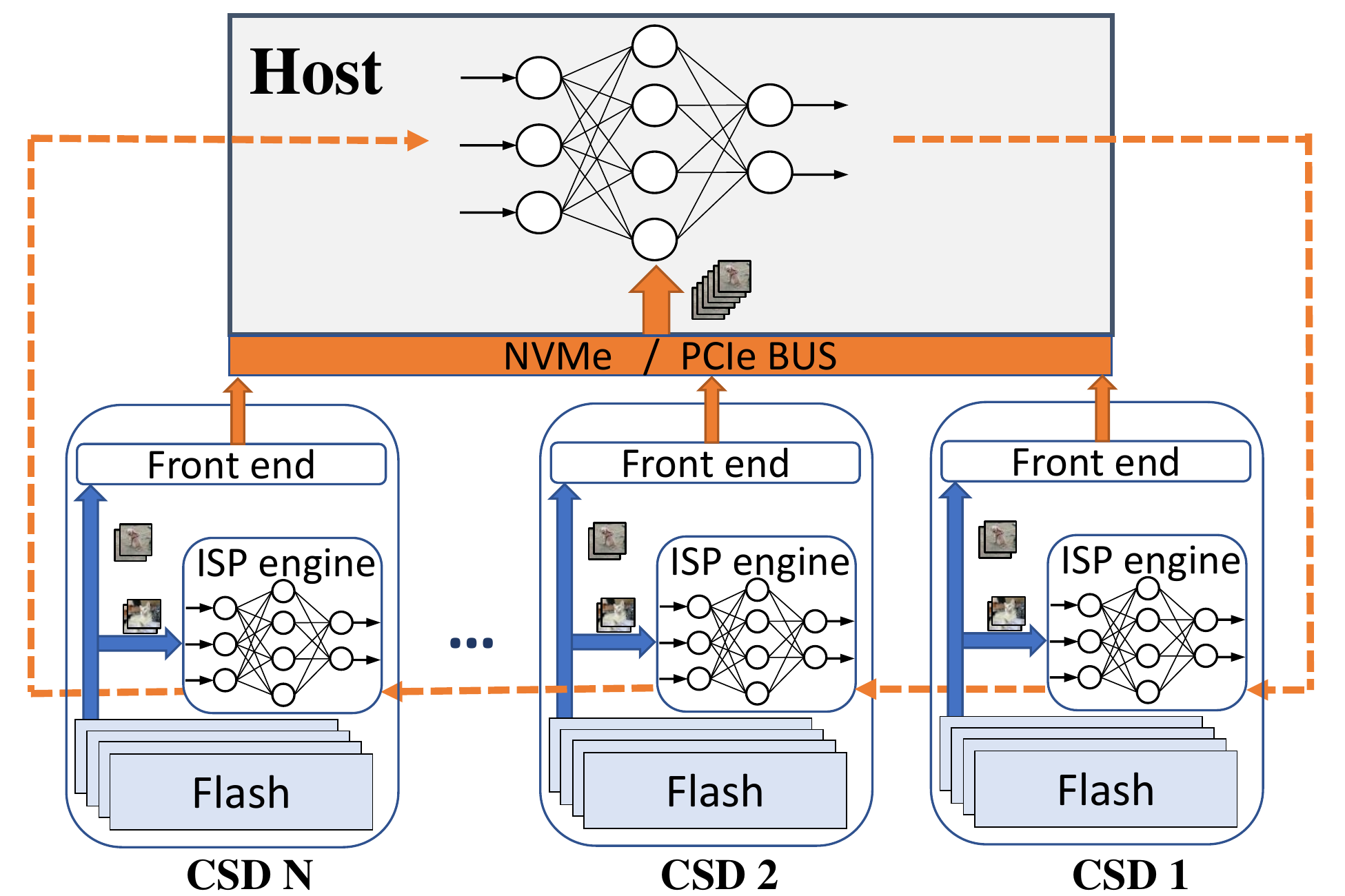}}
\caption{Stannis Architecture}
\label{fig_framework_arch}
\end{figure}
\begin{algorithm}[b]
\caption{Stannis's Tuning algorithm}
\label{alg:alg_1}
\KwIn{$\id{IP}_{\text{\Newport}},\id{IP}_{\text{host}},C$}
\KwOut{($\id{BS}_{\text{host}},\id{BS}_{\text{\Newport}}$)}
\SetKwFunction{FMain}{Tune}
\SetKwProg{Fn}{Function}{:}{}
\Fn{\FMain{$\id{IP}_{\text{\Newport}},\id{IP}_{\text{host}},C$}}
{
    \For{batch sizes in {list of BS}}{
    run $\id{benchmark}$ on {\Newport}\;
    update $\id{BS}_{\Newport}$  to the best one\;
    update $\id{time}_{\Newport}$\;
    }
    Let $E=$ margin scale \;
    \While{$(\id{time}_{\text{host}} – \id{time}_{\text{\Newport}}) < (\id{time}_{\text{\Newport}}/E)$}
    {
    $\id{BS}_{\text{host}} \text{+=} \id{BS}_{\text{host}}\times (\id{time}_{\text{\Newport}} – \id{time}_{\text{host}})/C$ \;
    run benchmark on host \;
    get the  $\id{time}_{\text{host}}$ \;
    }
    \Return $(\id{BS}_{\text{\Newport}}, \id{BS}_{\text{host}})$\;
}
\textbf{End Function}
\end{algorithm}
One potential concern with varying the training batch sizes is the loss of accuracy. In practice, however, we observed that the range of changes in batch size makes no tangible impact on accuracy.
While setting batch sizes can increase speed within one epoch, an imbalance in the datasets assigned to the computing nodes can stall faster nodes at the end of each epoch. After determining the optimal batch size, Stannis runs a load balancing algorithm to assign the proper number of input data to each node so all nodes finish after the same number of steps in one epoch. Since the host has access to more data than each individual CSD, it is convenient to determine the host's dataset size based on the batch size ratios and the slower node’s dataset size. The following equation shows how to determine dataset size on host:

\begin{equation}
\label{eq1}
\begin{split}
&\id{steps}_{\text{per epoch}}=\frac{\id{dataset}}{\id{batchsize}} \rightarrow \\
&\id{dataset}_{\text{host}}=\frac{\id{dataset}_{\text{card}}}{\id{batchsize}_{\text{card}}} \times \id{batchsize}_{\text{host}} 
\end{split}
\end{equation}

In case the portion of the private data to be processed on {\Newport} nodes are not equal, Stannis either uses more portion of the public data on the node with the smaller private database or duplicate the private data to maximize the rate of image per second.
Stannis can run any network that is based on Keras or Tensorflow. Examples of such are VGG, ResNet, Inception, MobileNetV2, and DenseNet. \Fig{fig_framework_arch} shows how the Stannis distributes and process data on system.

Distributing the training process means the central model receives updates less frequently that can cause accuracy loss. However, in \cite{ref_exp_facebook}, Goyal \textit{et al.} show that this accuracy degradation can be avoided to a good extent by applying two strategies: a) a linear scaling up of the learning rate based on the number of passed epochs, and b) a warm-up strategy that uses lower learning rates at the start of training, which helps overcoming early optimization difficulties \cite{ref_exp_resnet}.

\section{Experimental Results}\label{expr}
\label{sec:exp}

There are two scenarios for using Stannis in action. The first one is to run it on a server rack in a datacenter where the user data has been transferred to storage systems using end-to-end encryption to guarantee the privacy aspects. The second case considers {\Newport} as a stand-alone system, capable of doing edge computing in an IoT environment. Examples of such cases are the autonomous driving cars where lots of sensor data is created every second that can be processed fully or partially in the car. For now, we stick to the first case and explore the second one in our future work with the addition of the Federated learning techniques. 

\begin{figure}
\centerline{\includegraphics[scale=0.07]{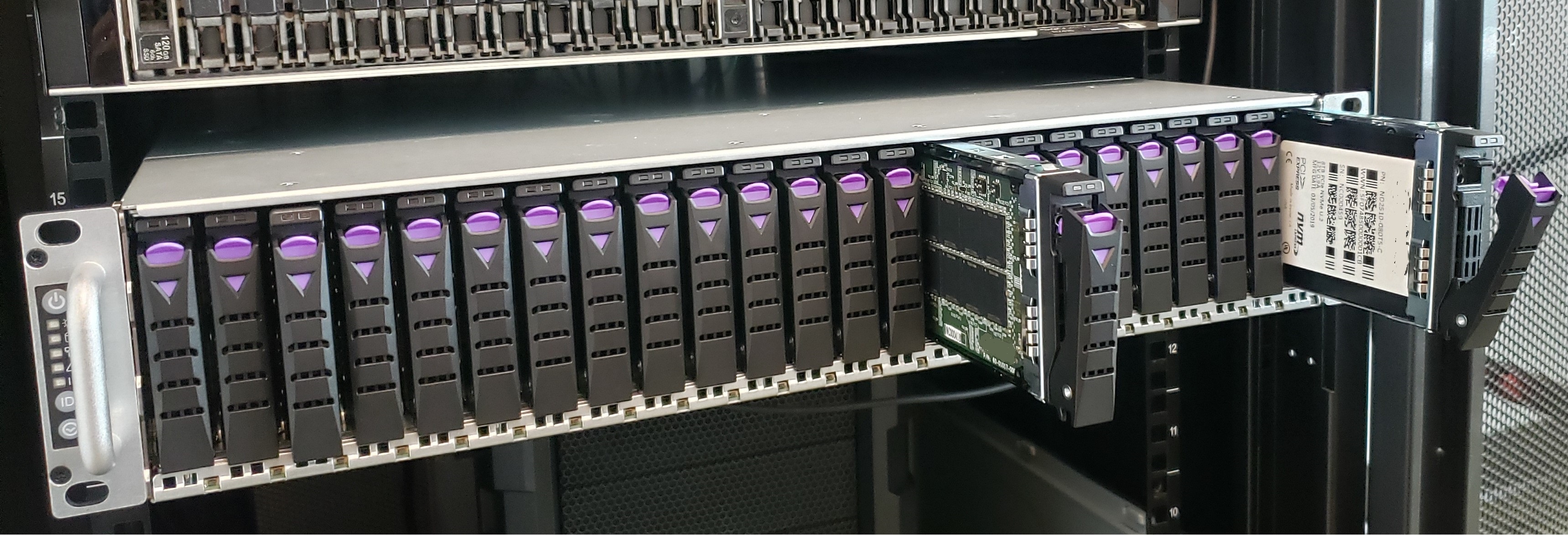}}
\caption{A server with 24 {\Newport} CSDs}
\label{fig_server}
\end{figure}

To evaluate Stannis in the real world, we assembled an AIC 2U-FB201-LX server with an 8-core, 16-thread Intel\textsuperscript{\textregistered} Xeon\textsuperscript{\textregistered} Silver 4108 CPU with 32GB of DRAM and 24 {\Newport} CSDs, each with 32~TB flash memory, giving us a storage server with a total of 24*32= 768~TB capacity (see \Fig{fig_server}). We used an expanded TinyImageNet database \cite{ref_exp_tinyimgnt} for neural network training and took 72000 images as public data and 12000 images as private distributed over all {\Newport} CSDs. We chose MobileNetV2 with 3.47 million parameters and 56 million multiply-and-accumulate (MAC) operations as our main neural network. 
To compare the speedup on different neural networks, we then ran the test for NASNet, InceptionV3, and SqueezeNet. The only concern in choosing a network and the batch size is the available DRAM on the systems. Large batch size on big networks can saturate the DRAM and thus stall the entire training process. The 6~GB DRAM on {\Newport} proved to be enough for most of the test cases. The solution for cases with DRAM saturation was to choose a smaller batch size. Since the processing speed converges to a certain number after a certain batch size, this reduction in batch size would not affect the processing speed. For instance, the images-per-second speed for MobilenetV2 on {\Newport} is about 3 images per second for all batch sizes greater than 16. This happens as a result of the full utilization of the processing engine when the task becomes computation intensive rather than communication intensive. 

\begin{table}
\caption{Parameter tuning from algorithm \ref{alg:alg_1}}
\resizebox{\columnwidth}{!}
{
\begin{tabular}{cccccc}
\hline
\textbf{Network}    
& \textbf{Param} 
& \textbf{Flop} 
& \textbf{MAC} 
& \textbf{\begin{tabular}[c]{@{}c@{}}batch size \\ Host / {\Newport}\end{tabular}} 
& \textbf{\begin{tabular}[c]{@{}c@{}}speed (img/sec) \\ Host / {\Newport}\end{tabular}}  \\ \hline

MobilenetV2 & 3.47M     & 7.16M     & 56M    & 315 / 25     &31.05 / 3.08\\
NASNet      & 5.3M      & 10.74M    & 564M    & 325 / 15    &47.31 / 2.80\\
InceptionV3 & 23.83M    & 47.82M    & 5.72G    & 370 / 16   &30.80 / 1.85\\
squeezenet  & 1.25M     &2.46M      & 861M    & 850 / 50    &219.0 / 16.3\\ \hline
\label{tab:table_1}
\end{tabular}
}
\end{table}
\begin{figure*}
\includegraphics[width=\textwidth]{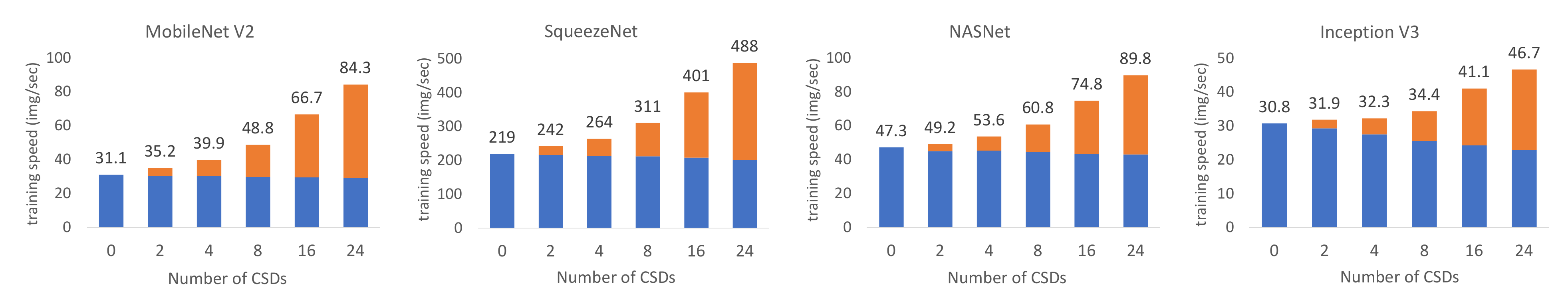}
\caption{Stannis Performance for different deep neural networks}
\label{speed1}
\end{figure*}
\begin{figure}
\centerline{\includegraphics[scale=0.4]{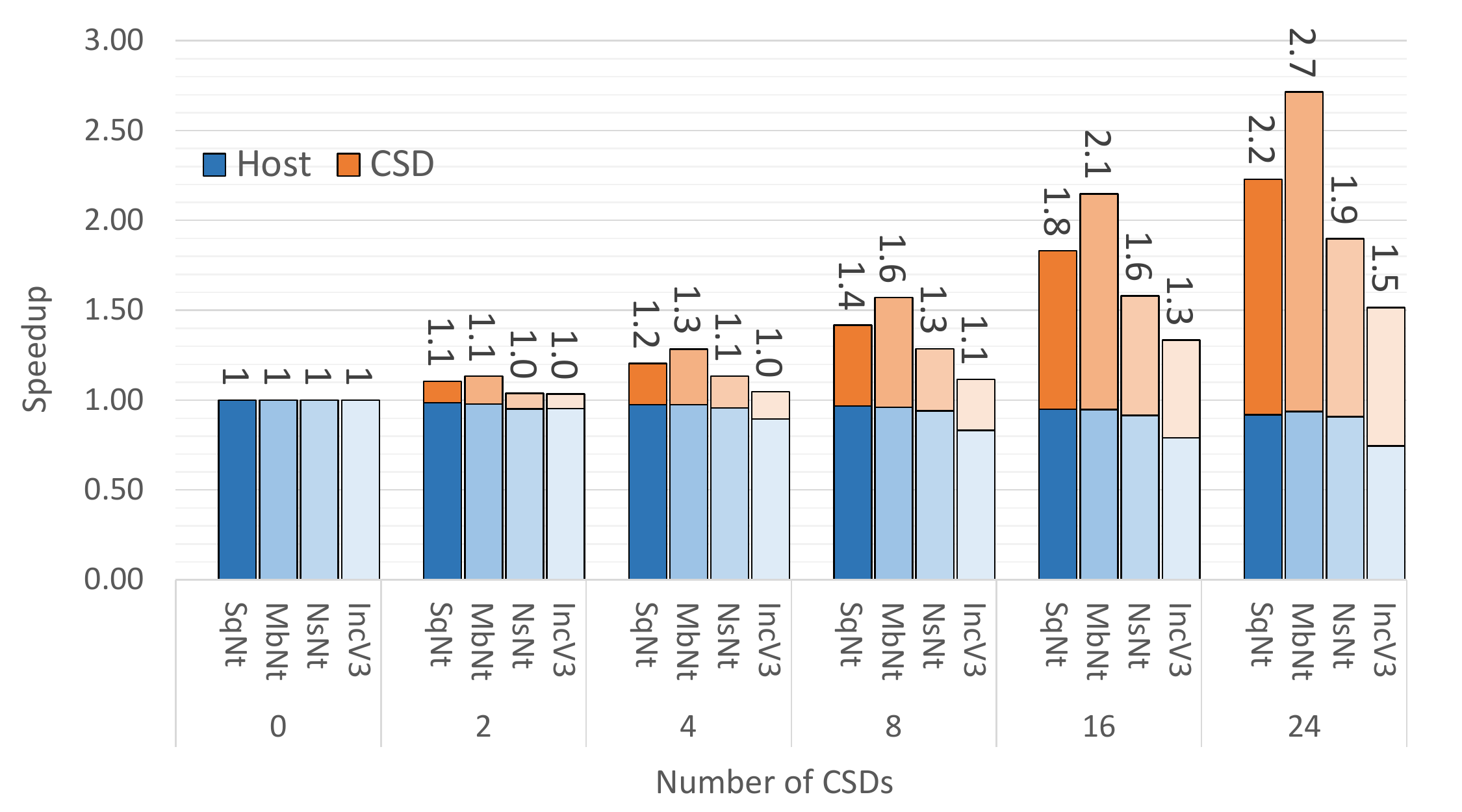}}
\caption{Speedup for different deep neural networks }
\label{speedgain}
\end{figure}

\subsection{Performance Analysis}

Stannis started with running the tuning algorithm for the MobileNetV2, which gave us the optimal batch sizes of 25 and 315 and the processing speed of 3 and 32.3 images per second for the {\Newport} and the host, respectively. It then determined the number of input data corresponding to each node to avoid stall on individual cards and then ran the benchmarks on a mixture of host and number of cards. We ran the same test for NASNet, InceptionV3, and SqueezeNet as well. The optimized batch sizes and corresponding speed for different networks are shown in Table~\ref{tab:table_1}. \Fig{speed1} shows the processing speed in the form of an image per second for different networks. There is a slowdown in all processing nodes in a distributed model, which is due to partial stalls when nodes are synchronizing the parameters. The slow down pace fades out, and the individual node’s performance converges to a certain speed after the number of nodes grows beyond 5-6 devices. The relative speedup for different networks is also shown in \Fig{speedgain}. The result shows that the smaller networks get better speedup than the larger ones; this is because the more parameters to update, the more time it takes to synchronize the nodes. Another important parameter is the number of MACs. As \Fig{speedgain} shows, SqueezeNet with 2.46M Flops gets less speedup compared to MobileNetV2 with 7.16M Flops, because it has 15x more MACs.

\subsection{Power Analysis}

The closest product to {\Newport} that we could find on the market was the 11~TB Micron MTFDHAL11TATCW-1AR1ZAB SSD. We evaluate the power consumption of an AIC server with 24 11-TB Micron SSDs against the same system with 24 32-TB {\Newport} CSDs. We used an off-the-shelf power meter to measure the input power to the entire server rack. Table~\ref{tab:energy_table} shows the energy per processed image for MobileNetV2. We skip the results for other networks since the power measurements were almost identical. Measurements show up to 69\% saving in energy per processed image and 2x FLOPS per watt with 24 {\Newport} CSDs compared to host alone.

\begin{table}
\caption{Energy consumption}
\resizebox{\columnwidth}{!}
{
\begin{tabular}{cccccc}\hline
\textbf{Number of CSDs}    & \textbf{0} & \textbf{4} & \textbf{8} & \textbf{16} & \textbf{24} \\ \hline
\textbf{Energy per image (J)}  & 13.10      & 8.30       & 6.84       & 5.05        & 4.02 \\
\textbf{Energy saving (\%)} & 0\%        & 37\%       & 48\%       & 62\%        & 69\% \\
\textbf{FLOPS per watt} & 5.87M        & 7.05M       & 8.18M       & 10.37M        & 12.26M \\ \hline   
\end{tabular}
}
\label{tab:energy_table}
\end{table}

\subsection{Accuracy Analysis}

To assess the accuracy degradation due to distribution of the training process, we ran a training session with 416,000 images on a single node and on six nodes to compare their accuracy and loss. The results show the loss increased from 1.1859 on a single node to 1.1907 on six nodes, or a 0.5\% increase, and the same accuracy of 0.31 on both setups.  Note that what is relevant here is not the absolute accuracy but the fact that our heterogeneous distributed setup yields the same accuracy as the centralized one.


\section{Conclusion}
\label{sec:concl}

This paper introduces a new in-storage processing solution to accelerate deep neural network training.  It achieves up to 2.7x speedup while reducing energy by up to 69\%.  Our solution consists of novel hardware named {\Newport}, a computational storage device (CSD) that augments processing power to the systems and eliminates the need to move data to the host for processing.  The hardware runs workload that is distributed by our software framework named Stannis, which extends Horovod to take advantage of heterogeneous processing units by optimizing the batch size and controlling the portion of data allotted to each processing node. Not only is this beneficial to both power and performance but more importantly it protects privacy by never moving private data out of the storage system. We plan to extend Stannis to a general framework that can parallelize any application on heterogeneous architectures and also develop a federated learning framework for training on mobile devices.

\bibliographystyle{IEEEtran}
\bibliography{reference}

\begin{thebibliography}{10}
\providecommand{\url}[1]{#1}
\csname url@samestyle\endcsname
\providecommand{\newblock}{\relax}
\providecommand{\bibinfo}[2]{#2}
\providecommand{\BIBentrySTDinterwordspacing}{\spaceskip=0pt\relax}
\providecommand{\BIBentryALTinterwordstretchfactor}{4}
\providecommand{\BIBentryALTinterwordspacing}{\spaceskip=\fontdimen2\font plus
\BIBentryALTinterwordstretchfactor\fontdimen3\font minus
  \fontdimen4\font\relax}
\providecommand{\BIBforeignlanguage}[2]{{%
\expandafter\ifx\csname l@#1\endcsname\relax
\typeout{** WARNING: IEEEtran.bst: No hyphenation pattern has been}%
\typeout{** loaded for the language `#1'. Using the pattern for}%
\typeout{** the default language instead.}%
\else
\language=\csname l@#1\endcsname
\fi
#2}}
\providecommand{\BIBdecl}{\relax}
\BIBdecl

\bibitem{ref_int_domo}
\BIBentryALTinterwordspacing
``Becoming a data-driven {CEO}.'' [Online]. Available:
  \url{https://www.domo.com/solution/data-never-sleeps-6}
\BIBentrySTDinterwordspacing

\bibitem{ref_int_AoI}
A.~Javani, M.~Zorgui, and Z.~Wang, ``Age of information in multiple sensing,''
  2019.

\bibitem{ref_int_cloud}
W.~A. Jansen, ``Cloud hooks: Security and privacy issues in cloud computing,''
  in \emph{2011 44th Hawaii International Conference on System Sciences}.\hskip
  1em plus 0.5em minus 0.4em\relax IEEE, 2011, pp. 1--10.

\bibitem{Torabzadeh_bigdata}
M.~Torabzadehkashi, S.~Rezaei, A.~HeydariGorji, H.~Bobarshad, V.~Alves, and
  N.~Bagherzadeh, ``Computational storage: an efficient and scalable platform
  for big data and hpc applications,'' \emph{Journal of Big Data}, vol.~6,
  no.~1, p. 100, 2019.

\bibitem{jun2015bluedbm}
S.-W. Jun, M.~Liu, S.~Lee, J.~Hicks, J.~Ankcorn, M.~King, S.~Xu \emph{et~al.},
  ``{Bluedbm}: An appliance for big data analytics,'' in \emph{2015 ACM/IEEE
  42nd Annual International Symposium on Computer Architecture (ISCA)}.\hskip
  1em plus 0.5em minus 0.4em\relax IEEE, 2015, pp. 1--13.

\bibitem{torabzadeh_HPCC}
M.~Torabzadehkashi, A.~Heydarigorji, S.~Rezaei, H.~Bobarshad, V.~Alves, and
  N.~Bagherzadeh, ``Accelerating hpc applications using computational storage
  devices,'' in \emph{21st International Conference on High Performance
  Computing and Communications}.\hskip 1em plus 0.5em minus 0.4em\relax IEEE,
  2019, pp. 1878--1885.

\bibitem{gu2016biscuit}
B.~Gu and et~al., ``{Biscuit}: A framework for near-data processing of big data
  workloads,'' in \emph{ACM SIGARCH Computer Architecture News}, vol.~44,
  no.~3.\hskip 1em plus 0.5em minus 0.4em\relax IEEE Press, 2016, pp. 153--165.

\bibitem{ref_bck_tensorflow}
Abadi \emph{et~al.}, ``{Tensorflow}: A system for large-scale machine
  learning,'' in \emph{12th USENIX Symposium on Operating Systems Design and
  Implementation OSDI}, 2016, pp. 265--283.

\bibitem{ref_bck_pytorch}
A.~Paszke, S.~Gross, S.~Chintala, G.~Chanan, E.~Yang, Z.~{DeVito}, Z.~Lin,
  A.~Desmaison, L.~Antiga, and A.~Lerer, ``Automatic differentiation in
  {PyTorch},'' in \emph{NIPS Autodiff Workshop}, 2017.

\bibitem{ref_bck_allreduce}
A.~Agarwal, O.~Chapelle, M.~Dud\'{i}k, and J.~Langford, ``A reliable effective
  terascale linear learning system,'' \emph{Journal of Machine Learning
  Research}, vol.~15, pp. 1111--1133, 2014.

\bibitem{ref_bck_horovod}
A.~Sergeev and M.~{Del Balso}, ``Horovod: fast and easy distributed deep
  learning in {TensorFlow},'' \emph{CoRR}, vol. abs/1802.05799, 2018.

\bibitem{ref_bck_ringallreduce}
\BIBentryALTinterwordspacing
``Bringing {HPC} techniques to deep learning.'' [Online]. Available:
  \url{http://andrew.gibiansky.com/}
\BIBentrySTDinterwordspacing

\bibitem{ref_bck_benchmark}
Y.~E. {Wang}, G.-Y. {Wei}, and D.~{Brooks}, ``{Benchmarking {TPU}, {GPU}, and
  {CPU} Platforms for Deep Learning},'' \emph{arXiv e-prints}, p.
  arXiv:1907.10701, Jul 2019.

\bibitem{rezaei_ICCD}
S.~Rezaei, K.~Kim, and E.~Bozorgzadeh, ``Scalable multi-queue data transfer
  scheme for fpga-based multi-accelerators,'' in \emph{36th International
  Conference on Computer Design (ICCD)}.\hskip 1em plus 0.5em minus 0.4em\relax
  IEEE, 2018, pp. 374--380.

\bibitem{kim2016storage}
S.~Kim, H.~Oh, C.~Park, S.~Cho, S.-W. Lee, and B.~Moon, ``In-storage processing
  of database scans and joins,'' \emph{Information Sciences}, vol. 327, pp.
  183--200, 2016.

\bibitem{park2016storage}
D.~Park, J.~Wang, and Y.-S. Kee, ``In-storage computing for {Hadoop}
  {MapReduce} framework: Challenges and possibilities,'' \emph{IEEE
  Transactions on Computers}, 2016.

\bibitem{torabzadeh_PDP}
M.~Torabzadehkashi, S.~Rezaei, A.~Heydarigorji, H.~Bobarshad, V.~Alves, and
  N.~Bagherzadeh, ``Catalina: in-storage processing acceleration for scalable
  big data analytics,'' in \emph{27th Euromicro International Conference on
  Parallel, Distributed and Network-Based Processing (PDP)}.\hskip 1em plus
  0.5em minus 0.4em\relax IEEE, 2019, pp. 430--437.

\bibitem{ref_rltd_compstor}
M.~Torabzadehkashi, S.~Rezaei, V.~Alves, and N.~Bagherzadeh, ``Compstor: An
  in-storage computation platform for scalable distributed processing,'' in
  \emph{International Parallel and Distributed Processing Symposium Workshops
  (IPDPSW)}.\hskip 1em plus 0.5em minus 0.4em\relax IEEE, 2018, pp. 1260--1267.

\bibitem{ref_exp_facebook}
\BIBentryALTinterwordspacing
P.~Goyal, P.~Doll{\'{a}}r, R.~B. Girshick, P.~Noordhuis, L.~Wesolowski,
  A.~Kyrola, A.~Tulloch, Y.~Jia, and K.~He, ``Accurate, large minibatch {SGD:}
  training {ImageNet} in 1 hour,'' \emph{CoRR}, vol. abs/1706.02677, 2017.
  [Online]. Available: \url{http://arxiv.org/abs/1706.02677}
\BIBentrySTDinterwordspacing

\bibitem{ref_exp_resnet}
\BIBentryALTinterwordspacing
K.~He, X.~Zhang, S.~Ren, and J.~Sun, ``Deep residual learning for image
  recognition,'' \emph{CoRR}, vol. abs/1512.03385, 2015. [Online]. Available:
  \url{http://arxiv.org/abs/1512.03385}
\BIBentrySTDinterwordspacing

\bibitem{ref_exp_tinyimgnt}
\BIBentryALTinterwordspacing
``Tiny imagenet visual recognition challenge.'' [Online]. Available:
  \url{https://tiny-imagenet.herokuapp.com/}
\BIBentrySTDinterwordspacing

\end{thebibliography}

\end{document}